 \documentclass[12pt,preprint,usenatbib]{aastex}  

\def\dsm{$\mathrm{M}_\odot$}

\shorttitle{BINARY AND ROTATION FOR EXPLAINING CMDS} \shortauthors{Li Z.M. et al.}

\begin{document}


\title{Combined Effects of Binaries and Stellar Rotation on the Color-Magnitude Diagrams of Intermediate-Age Star Clusters}

\author{Zhongmu Li\altaffilmark{1,2}, Caiyan Mao\altaffilmark{1}, Li Chen\altaffilmark{1}, Qian Zhang\altaffilmark{1}}

\altaffiltext{1}{Institute for Astronomy and History of Science and
Technology, Dali University, Dali 671003, China;
zhongmu.li@gmail.com} \altaffiltext{2}{National Astronomical
Observatories, Beijing 100012, China}

\begin{abstract}
About seventy percent of intermediate-age star clusters in the Large
Magellanic Clouds have been confirmed to have broad main sequence,
multiple or extended turn-offs and dual red giant clumps. The
observed result seems against the classical idea that such clusters
are simple stellar populations. Although many models have been used
for explaining the results via factors such as prolonged star
formation history, metallicity spread, differential redenning,
selection effect, observational uncertainty, stellar rotation, and
binary interaction, the reason for the special color-magnitude
diagrams is still uncertain. We revisit this question via the
combination of stellar rotation and binary effects. As a result, it
shows ``golf club'' color-magnitude diagrams with broad or multiple
turn-offs, dual red clump, blue stragglers, red stragglers, and
extended main sequences. Because both binaries and massive rotators
are common, our result suggests that most color-magnitude diagrams
including extended turn-off or multiple turn-offs can be explained
using simple stellar populations including both binary and stellar
rotation effects, or composite populations with two components.
\end{abstract}


\section{Introduction}

Star clusters are usually assumed to be simple stellar populations (SSPs) that
all stars formed in a short timescale and with the same metallicity.
The color-magnitude diagram (CMD) of a star cluster should be similar to an SSP isochrone.
However, recent observations based on the data that are obtained with
the Advanced Camera for Surveys (ACS) onboard the Hubble Space
Telescope (HST) showed double, multiple or extended main sequence (MS) turn-offs (MSTOs),
and dual clump of red giants (RC) for many intermediate-age star clusters in the Large Magellanic Cloud (LMC),
e.g., NGC1751, NGC 2108, NGC 1846, NGC1806, and NGC1783.
The shape of CMDs near turn-off looks like a golf club (hereafter ``golf club'' shape).
One can refer to many papers such as \cite{bert03}, \cite{piot05}, \cite{piot07},
\cite{milo09,milo10}, \cite{mack07, mack08, milo09}, \cite{goud09}, \cite{bast09}, \cite{Girardi11},
and \cite{Rubele11} for more details.

Different models have been used for interpreting the special CMDs.
For example, \cite{mack08} and \cite{goud09} explained this by
spread of chemical abundance, but it is disagreement with the
similarity of metallicity of all stars
\citep{mucc08,Goudfrooij11b,Goudfrooij11a}. Then \cite{mack08,
goud09} interpreted this by capture of field stars. This reproduces
extended MSTO but has difficulty for explaining double or multiple
MSTOs (MMSTOs), which was found by e.g., \cite{glat08}. Similarly, a
picture based on merger of existing star clusters had been brought
forward \citep{mack07}, but it seems not common that clusters in a
normal molecular cloud merge with each other \citep{goud09}. A
scenario of formation of a second generation of stars from the
ejecta of first generation asymptotic giant branch stars was
suggested but it also has some disadvantages (\citealt{derc08} and
\citealt{goud09}). \cite{bast09} studied this problem using the
effect of rotation on the effective and surface gravity of stars,
and then on the color and magnitude of star clusters. This model
reproduced the double or multiple population, but the work of
\cite{rube10} and \cite{Girardi11} argued that the effect of stellar
rotation alone could not explain the presence of ``golf club'' shape
of MSTOs. Instead, they suggested a prolonged star formation history
to explain the CMD with ``golf club'' shape \citep{Girardi11}. An
age spread about 300 Myr is needed for explain ``golf club'' CMDs
including both MMSTOs and double RC in such picture. However, this
brings a new challenge for the present dynamical formation model of
stars and star clusters. Some works also studied the possibility of
using observational selection \citep{kell11} and uncertainty to
interpret the observed CMDs but it seems not natural
\citep{Goudfrooij11a}. The kind of model involving a mixture of
stars with and without overshooting can not fit the observed CMD,
either \citep{Girardi11}. A latest model \citep{platais12} showed
the potential of differential reddening for explaining the apparent
splitting/widening of MSTOs of an Galactic open cluster, Trumpler
20, but an age spread is needed. Although most works give their
explanation based on single stars, a few works argue that unresolved
binary stars may result in extended MS and dual RC
\citep{milo09,Yang11}, but it can not reproduce some peculiar CMD
structures, e.g., the ``golf club'' shape \citep{mack08, goud09,
milo09}. Even so, the importance of stellar rotation and binaries in
modeling CMDs is obvious (\citealt{Hurley98}, \citealt{goud09},
\citealt{bast09}, \citealt{Li2010}, \citealt{Girardi11} and
\citealt{Yang11}). Because it is well known that all star clusters
possibly contain a large number of binaries (e.g.,
\citealt{Abt1979,Lada2006}) and massive rotating stars (e.g.,
\citealt{Peterson2004,McAlister2005,royer07}), it is necessary to
study the CMDs of intermediate-age star clusters via a combination
of the two natural factors in more detail. This paper just aims to
do this. The difference between this work and previous ones (e.g.,
\citealt{Girardi11,Yang11}) is on the construction of stellar
populations and treatment of rotational effect. The effect of
stellar rotation is considered using the result of \cite{bast09} and
taking different distribution for rotation rate ($\omega$). The
treatment of stellar rotation of \cite{bast09} is actually different
from \cite{Girardi11}, but it agrees with the results of Geneva
group \citep{bast09}. Stars with different masses are considered in
our model and a binary fraction similar to the observed one is used,
which makes the theoretical population more similar to real star
clusters compared to \cite{Yang11}. The results are finally found to
be different from previous works.

The structure of this paper is as follows. Section 2 introduces the
construction of CMDs. Then Section 3 shows the main results.
Finally, we summarize and discuss on the work in Section 4.

\section{Construction of CMDs}
We construct synthetic CMDs flowing our previous work (e.g.,
\citealt{Li2008mn,Li2008apj,Li2008iaus,Li2010,Li2011,Li2012}) on
modeling binary star stellar population (bsSP) and single star
stellar population (ssSP). In detail, the initial mass function
(IMF) of \cite{Salpeter1955} is adopted to generate our sample
stars. In order to build up binaries, the mass of primary component
of a binary is generated first within the range from 0.1 to 100
\dsm{}. Note that we do not exclude any stars like \cite{Yang11}, as
two components in an interactive binary can transfer and change
their masses in their evolution. Then the mass of the secondary
component is calculated via taking a random secondary-to-primary
mass ratio ($q$) \citep{han95}. $q$ is assumed to obey an uniform
distribution within 0--1. The separations ($a$) of two components is
given under the assumption that $a$ is constant in $\log a$ for wide
binaries and falls off smoothly at close separation \citep{han95},
which can be expressed by:
\begin{equation}
an(a)=\left\{
 \begin{array}{lc}
 \alpha_{\rm sep}(a/a_{\rm 0})^{\rm m} & a\leq a_{\rm 0};\\
\alpha_{\rm sep}, & a_{\rm 0}<a<a_{\rm 1},\\
\end{array}\right.
\end{equation}
where $\alpha_{\rm sep}\approx0.070$, $a_{\rm 0}=10R_{\odot}$,
$a_{\rm 1}=5.75\times 10^{\rm 6}R_{\odot}=0.13{\rm pc}$ and
$m\approx1.2$. This gives about half binary stars with orbital
periods less than 100 yr \citep{han95} and half single stars. After
that, a random eccentricity (e) within 0-1 is assigned to each
binary, as $e$ affects the evolution of stars slightly
\citep{Hurley02}. Because the binary fraction of real clusters is
possibly lower than 50\% (about 30--40\%, see e.g.,
\citealt{elson98}), we define our sample by removing some random
binaries from the generated sample. It leads to a new sample with
binary fraction of 35\%, which is then used in this work.

After the sample generation, all stars are evolved using the rapid
stellar evolution code of \cite{Hurley02} (Hurley code), which uses
some formulae fitted from evolutionary tracks of stellar models to
evolve stars and does not take stellar rotation into account. For
stellar populations including rotational stars, we add the effect of
rotation on effective temperature and luminosity to the parameters
of massive ($>$ 1.2 \dsm{}) stars that have not left main sequence
(MS), using two fitted correlations presented by \cite{bast09}. This
treatment is followed the work of \cite{bast09}. However, we adopt a
Gaussian distribution with mean and standard deviation of 0.55 and
0.25 (or 0.15) for rotation rate ($\omega$), according to the
$\omega$ distribution of some A and F type stars \cite{royer07} and
similar application by \cite{bast09}. We also assume that $\omega$
increases gradually with stellar mass, which gives the final
$\omega$ by multiplying a factor within 0 to 10 for each star (see
also \citealt{bast09}). The factor is linearly increase with mass,
and the values for 1.2 \dsm{} and 1.5 \dsm{} are zero and one
respectively. The effect of inclination is randomly set to about 0.2
times of that caused by $\omega$, according to the result of
\cite{bast09}. A small blue shift of about 0.01\,mag in $(V-I)$ of
fast rotators is also taken into account, according to the result of
\cite{platais12}. Some limitations in the treatment of stellar
rotation will be discussed later. Finally, we transform the
evolutionary parameters ([Fe/H], T$_{eff}$, $\log g$, $\log L$) into
colors and magnitudes using the atmosphere library of \cite{leje98}.

\section{Results}
Figure 1 shows the CMDs of a few binary star simple stellar
populations (bsSSPs) and binary star composite stellar populations
(bsCSPs) with $Z$ = 0.008. The CMDs in six panels are generated by
taking different assumptions, and the last four are similar to
observed CMDs of intermediate-age star clusters in LMC.

Panel $(a)$ shows the CMD of a bsSSP when all stars could be
resolved. It aims to show the direct effect of binary evolution or
interactions on CMD. A binary fraction of 35\% and star number of
76\,922 are taken for this population. We see that binary
interactions lead to some blue stragglers (BSs) and dual RC. BSs are
mainly caused by mass transfer, star merger, and possibly high
runaway velocity of stars (see also \citealt{Pols1994}). The dual RC
results from both normal stars (single and unmerged binary
components), and merged (most) or interactive binaries. In addition,
the turn-off is obviously dominated by single stars in the
population. This implies that the observed turn-off spread is
contributed from binary interaction slightly. Note that the turn-off
mass in the population is about 1.5 \dsm and the deep sequence is
close to the isochrone of an ssSSP with the same metallicity and
age. Furthermore, some red stragglers (or subgiants) (hereafter RSs)
are shown on the right of MS and under the giant branch. They are
primary binary components that transferred mass to secondary ones.
Some of observed stars in clusters (e.g., NGC 1846) are possibly
such stars.

Then panel $(b)$ gives a CMD by taking the spatial resolution of HST
ACS into account. The distance of LMC is taken as 160\,000
light-year, and the angle between line-of-sight and the connecting
line of two binary components is given randomly. This results in
about 85\% unresolved binaries. We find that MS becomes wider and
two parts (blue and red) are shown, in which the blue one is
obviously dense and narrow. At the same time, double turn-offs (2
MSTOs) are presented, but their locations seems different from CMDs
of star clusters in LMC. Comparing to panel $(a)$, we know that the
upper one of two MSTOs is caused by unresolved binaries. For
convenience, we cite the effect of both binary evolution and
resolution as binary effect in the paper. Note that the CMD does not
have ``golf club'' shape.

Next, in panel $(c)$ we consider both the effects of binaries and
stellar rotation in a population. When the effect of rotation is
taken into account, the population shows a special ``golf club''
shape similar to the observed CMDs of some LMC clusters. In detail,
the bsSSP with rotational stars has obvious spread in main sequence
and extended turn-off. It also shows dual RC and BSs. As a test, we
find that the number (about 7) of BSs in a fixed range, which is
shown by blue lines and affected by field star slightly, is similar
to that in NGC 1846 when considering only quality-filtered stars
within 30" from the center (see figure 3 of \citealt{mack08} for
comparison). The distance modulus and color excess are taken as
18.45 and 0.1\,mag respectively in the comparison. The panel shows
that single stars, separated binary components, and unresolved stars
have some contributions to the shape of CMD.

When comparing the CMD with an isochrone (red points in panel $c$)
of a single star simple stellar population (ssSSP) with the same
metallicity and age as the bsSSP, we find that the ssSSP isochrone
locates near the blue and dense MS part of bsSSP. It implies that
stellar metallicity and age can possibly be estimated by fitting the
dense MS using isochrones of ssSSPs. Because the turn-off part lower
than the ssSSP isochrone is mainly caused by stellar rotation, it
does not indicate more populations. However, as shown by
\cite{Girardi11}, it can be explained using prolonged star formation
history.

In panel $(d)$, we show the CMD for the same population, but with a
narrow distribution of rotation rate for some stars. In this test,
half stars are assume to rotate with a Gaussian distribution for
$\omega$, which peaks at 0.55, and with a standard deviation of
0.15. We take this assumption because some works show bimodal
$\omega$ distributions that peak near 0.1 and 0.5-0.6 for some A and
F type stars (e.g., \citealt{royer07}). Our treatment gives similar
results as taking two peaks of 0.1 and 0.55 for $\omega$. This model
reproduces the CMD with double turn-offs. Similar CMDs have been
observed in some star clusters. Because some other works used two
ssSSPs to explain similar CMDs, we plot the isochrones of two ssSSPs
with ages of 1.4 and 1.7\,Gyr in this figure as a comparison. We can
see that rotation effect can also be interpreted as one more stellar
population.

By comparing CMDs in panels $(c)$ and $(d)$ to those of star
clusters we find that they can fit to the observed CMDs of most
intermediate-age star clusters. Besides the extended MS or MMSTOs,
bsSP with rotation effect can fit the BSs, dual RC, and some RSs
(e.g., NGC 1846 and NGC 1987) (see, e.g., \citealt{milo09})
naturally. Note that RSs can be reproduced by assuming more total
and resolved binaries. Therefore, it is possible that many star
clusters are actually bsSSPs with stellar rotation. This supports to
the classical picture of stellar population of star clusters, i.e.,
the SSP scenario.

As some works argued that prolonged star formation history may be
the best choice for interpreting the CMDs of star clusters (e.g.,
\citealt{Girardi11}), it is necessary to compare the CMD of bsSSP to
that of composite stellar populations (CSPs). We do this by panels
$(e)$ and $(f)$. Panel $(e)$ gives a CMD of a bsCSP with the same
metallicity as the bsSSP in panel $(c)$. The bsCSP is assumed to
form its stars within 300\,Myr from 1.5\,Gyr by four star bursts. We
see that this case reproduces wider MSTO and RC. It also has a
``golf club'' shape. Similarly, panel $(f)$ shows a bsCSP with two
subpopulations, and considering the effect of a narrow $\omega$
distribution. We find that more MSTOs are generated. This suggests
that the picture of continuous star formation history is not the
only one for explaining MMSTOs (see \citealt{Girardi11} for
comparison). A simple model with two star bursts or a merger of two
clusters can potentially explain most CMDs with MMSTOs.

\section{Conclusion and Discussion}
This paper uses the combined effects of binaries and stellar
rotation to interpret the CMDs of intermediate-age star clusters in
LMC. It is shown that most observed CMD features can be reproduced
via both bsSSP or bsCSP with two subpopulations. In detail, the
model can show MMSTOs, broad main sequences with two parts, dual RC,
and blue and red stragglers. It suggests that MS spread and MMSTOs
possibly result from both rotation and binary effect, while dual RC,
blue and red stragglers are mainly caused by binary interactions
(merger and mass transfer). In addition, the combined effects of
binaries and stellar rotation leads to CMDs with a ``golf club''
shape turn-off, which is similar to the observed result.
Furthermore, bsSPs are shown to have some intrinsic spread in MS,
which results from unresolved binaries. By taking different
assumptions for stellar rotation rate, both extended or double MSTO
can be reproduced from a bsSSP. Therefore, the special CMD shapes of
many intermediate-age star clusters in LMC can be explained by
including both stellar rotation and binary effects in SSP models. In
this case, many star clusters with special CMDs are possibly SSPs,
rather than CSPs. Even if CSPs are needed, some models with two
components can potentially explain the observed CMDs.

Although the simple model considering both binary effect and stellar
rotation seems successful to reproduce the shape of CMDs of
intermediate-age star clusters, it is far from well understanding
the observed CMDs. Firstly, there are many uncertainties in the
model. The density of blue and red MS parts, the numbers of blue and
red stragglers are directly relating to the numbers of total and
unresolved binaries. Besides, while the treatment of rotation is
derived from detailed stellar evolutionary models, the effects of
rotation on stellar interiors and there appearance is considerably
uncertain. It should be noted that these models are 1D models in
which rotation is treated in a simple diffusive approximation. On
the more massive end of the spectra serious questions have been
raised about the effects of rotation on interior mixing
\citep{Hunter2011,Brott2011}. Secondly, the difference between the
results of this work and \cite{Girardi11} mainly results from
various treatment of stellar rotation, because different assumptions
(single or distributed) for rotation rate can lead to quite
different CMD features. Although it is reasonable to take a Gaussian
distribution for rotation rate following the observational result,
and assume that massive stars rotate faster since lower mass stars
have convective envelopes, which may generate magnetic fields
through dynamos spinning the stars down, our treatment is actually
simplistic and artificial. This needs to be done better in future
studies using either an empirical mass dependence or a grid or
detailed models accounting for rotation and a certain magnetic
braking prescription. In addition, two simple fitting formulae were
used for modeling the effect of stellar rotation, so the effect of
rotation on stars' lifetime, which is mainly responsible to
different color shifts caused by rotation in this work and
\cite{Girardi11}, and the relation between binarity and rotation
have not been taken into account. It should be better to use a
stellar evolutionary code including both binary and rotation in
future works. Furthermore, The effects of binary interaction and
rotation are implemented as two independent effects. In fact,
binarity may in many cases be the cause of rapid rotation. Moreover,
this work did not consider the dynamical and chemical evolution of
star clusters. They may supply important clues for better
understanding some gradients for building CMDs, especially star
formation history. Finally, many factors may affect the observed
CMDs, and the roles of them (e.g., prolonged star formation and
rotation) are somewhat degenerate. This can only be disentangled by
a series of works involving both the improvement in observation and
theoretical modeling.

\acknowledgments  We thank the referee for constructive comments and
Eva K. Grebel for suggestion. This work has been supported by the
Chinese National Science Foundation (Grant Nos. 10963001, 11203005),
Yunnan Science Foundation (No. 2009CD093), and Chinese Postdoctoral
Science Foundation. ZL gratefully acknowledges the support of
Sino-German Center (GZ585) and K. C. Wong  Education Foundation,
Hong Kong.

\begin{figure} 
\centering
\includegraphics[angle=-90,width=0.45\textwidth]{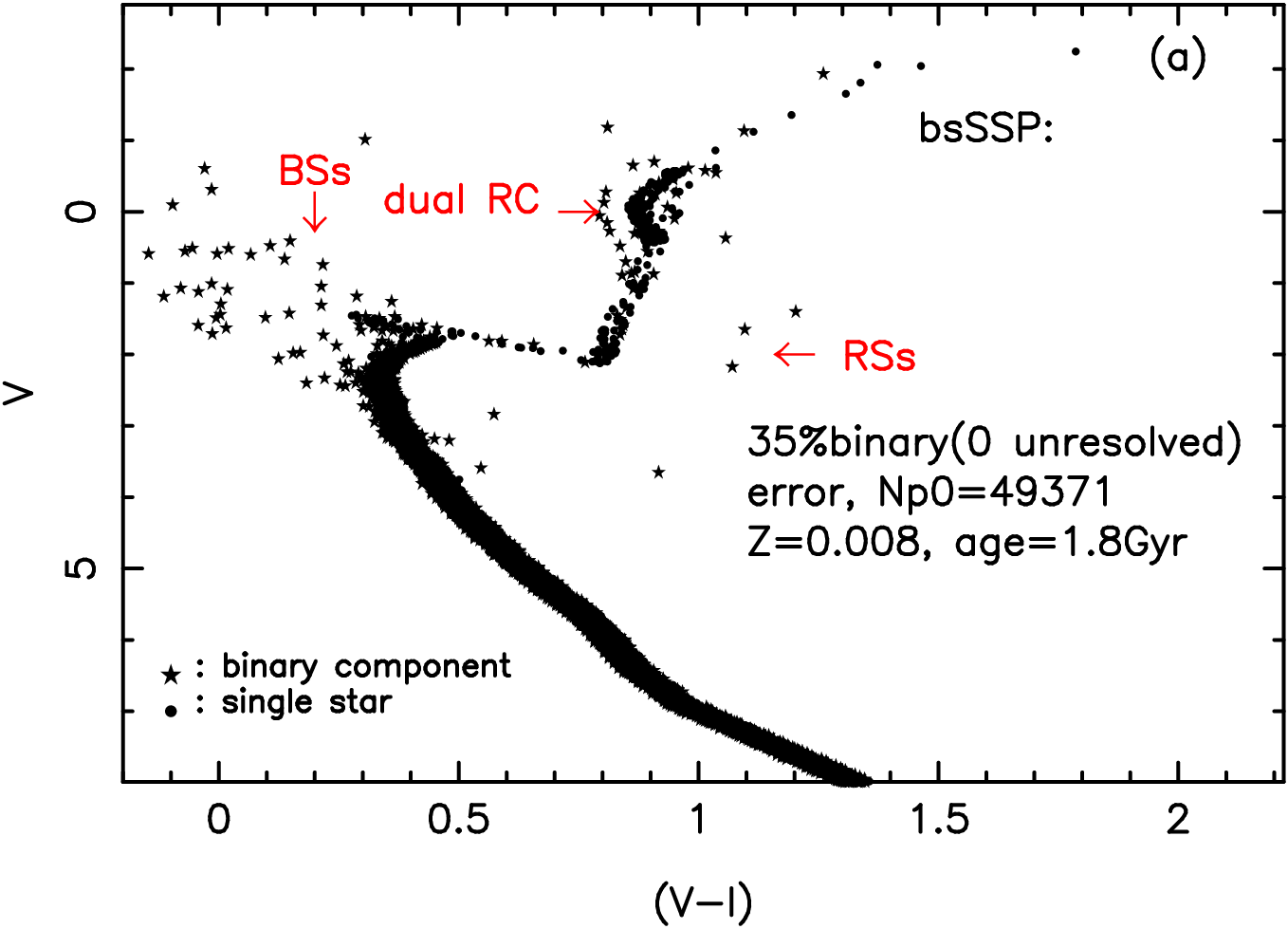}
\includegraphics[angle=-90,width=0.45\textwidth]{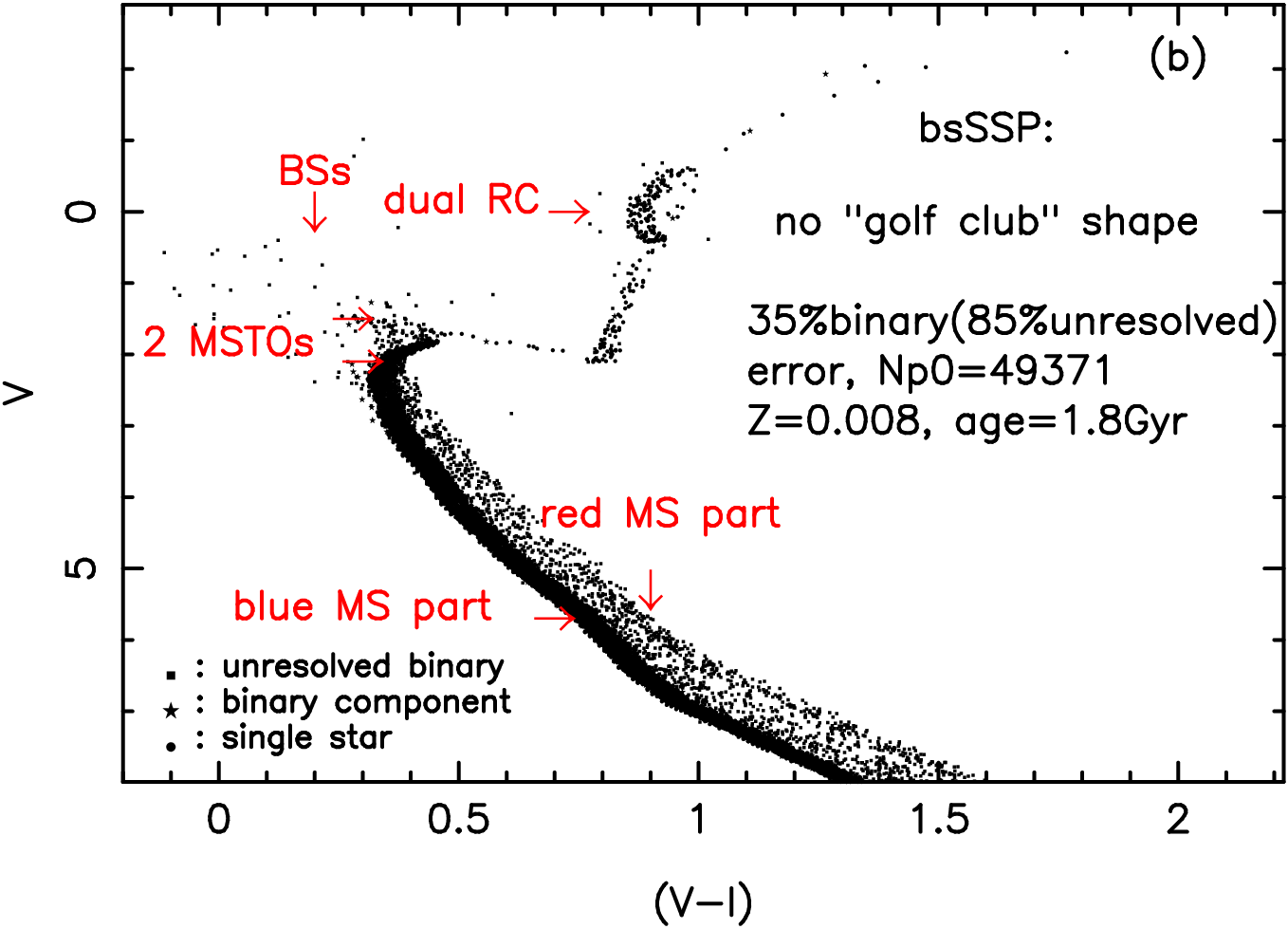}
\includegraphics[angle=-90,width=0.45\textwidth]{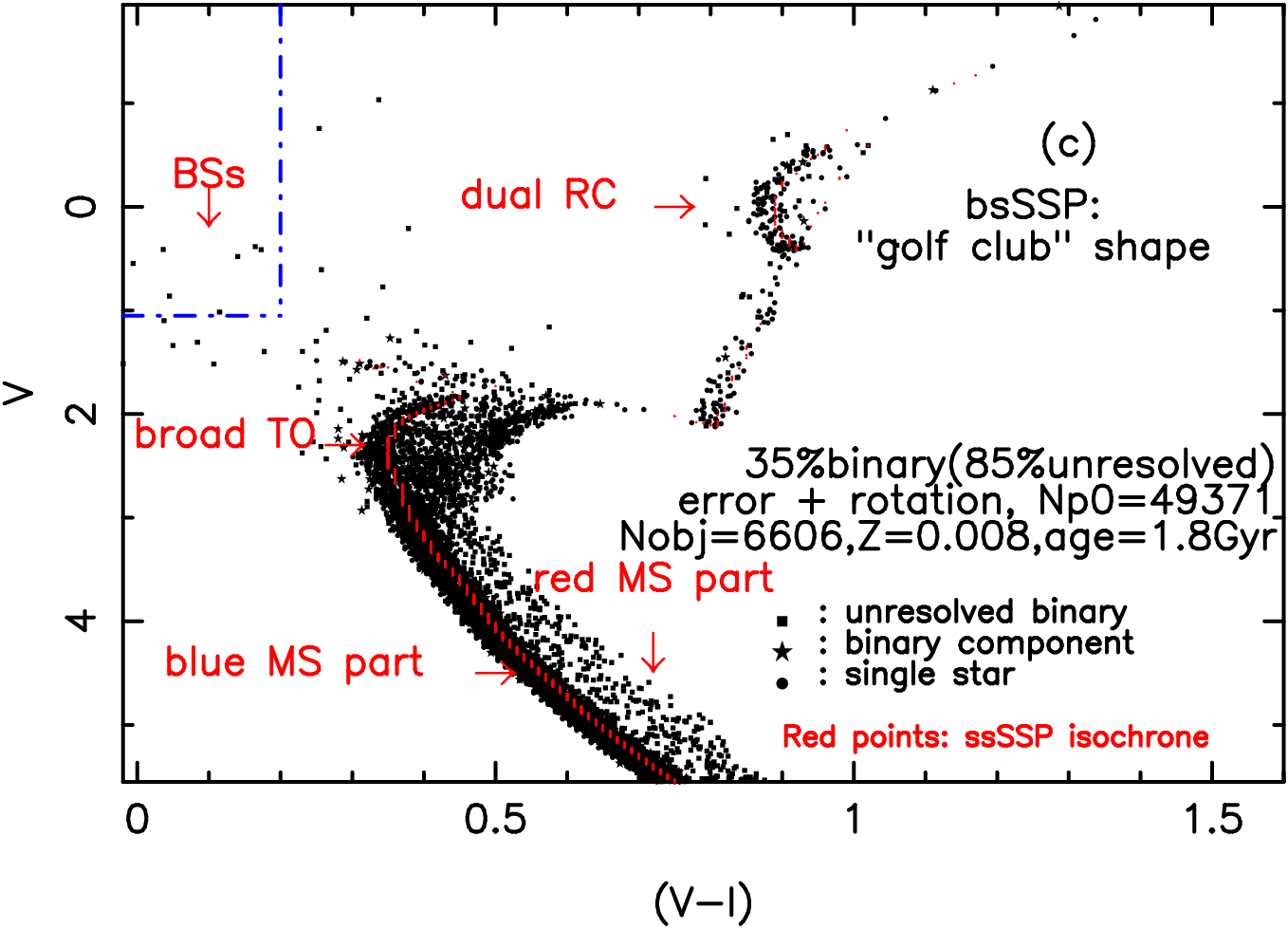}
\includegraphics[angle=-90,width=0.45\textwidth]{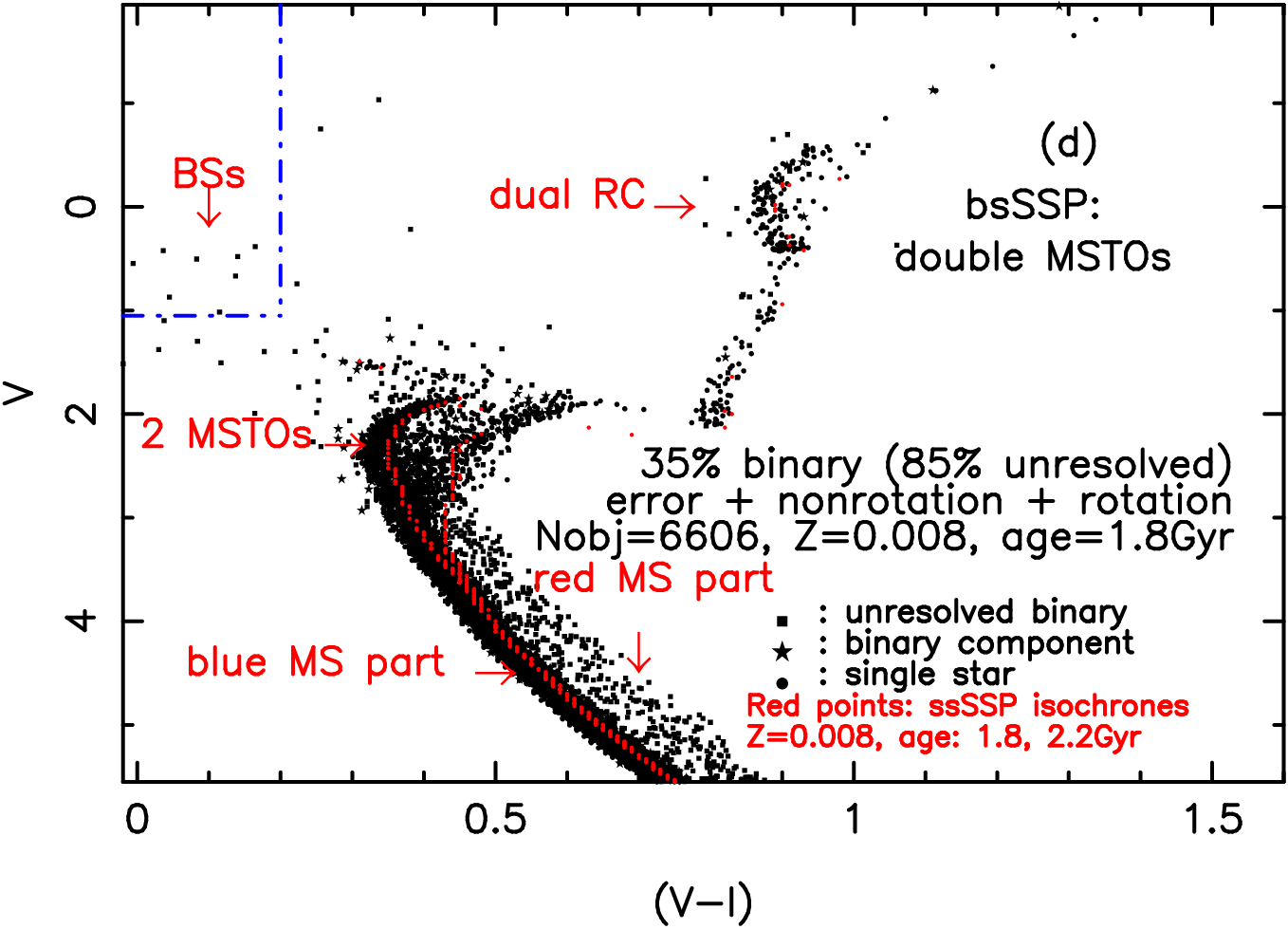}
\includegraphics[angle=-90,width=0.45\textwidth]{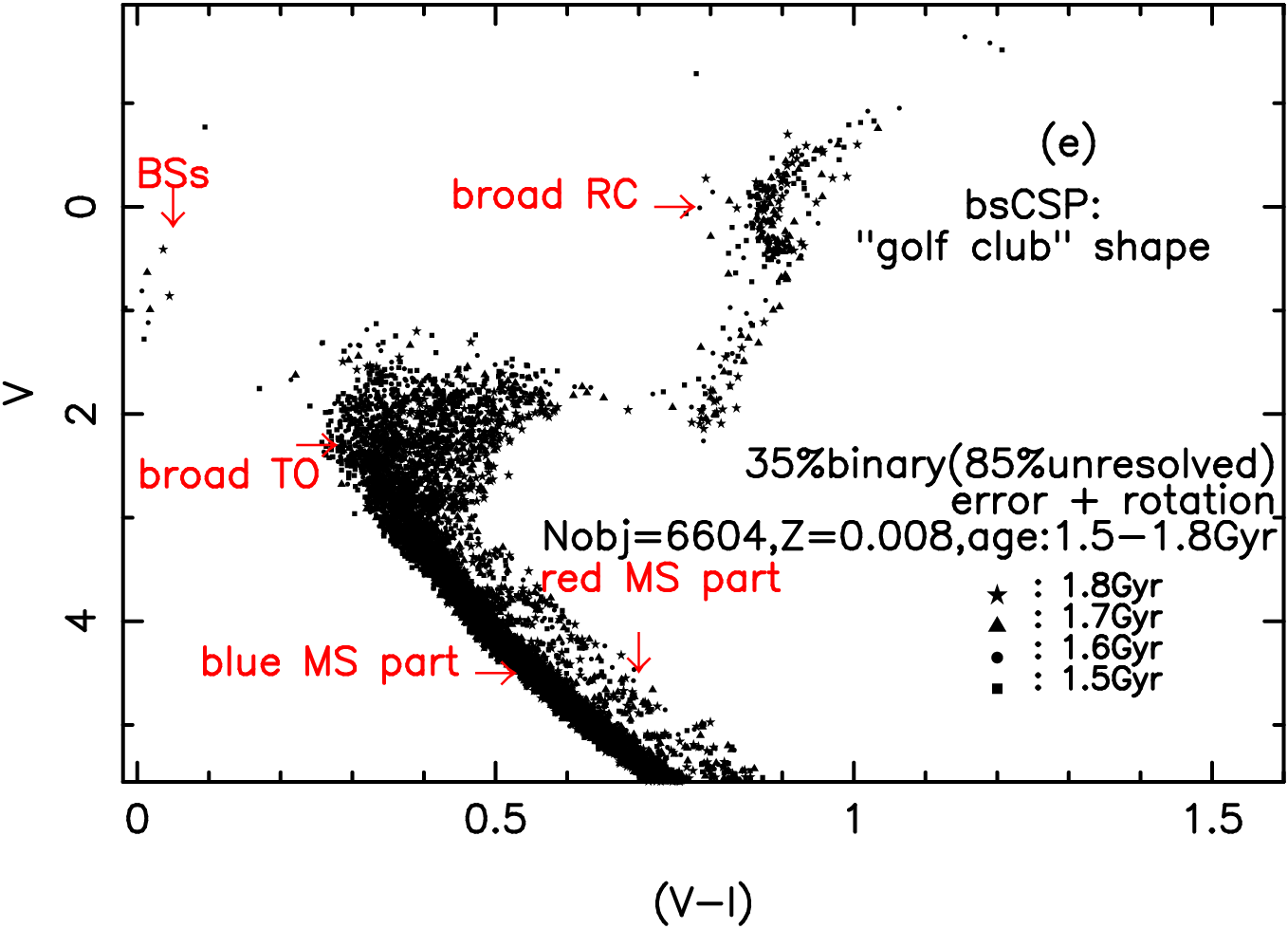}
\includegraphics[angle=-90,width=0.45\textwidth]{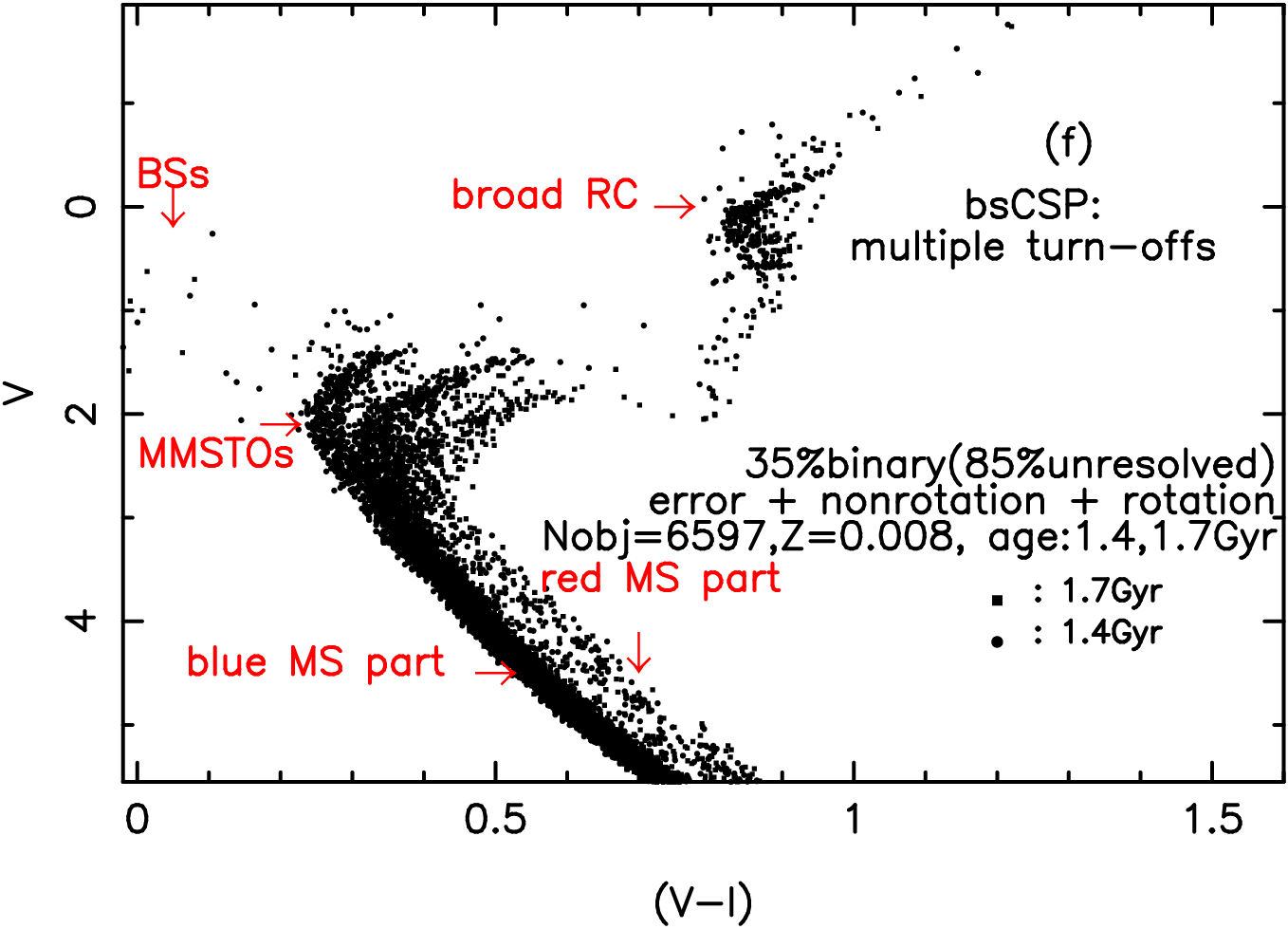}
\caption{Synthetic CMDs.
``bsSSP'' and ``bsCSP'' denote binary star simple and composite stellar population, respectively.
Fraction of unresolved binaries is calculated from binary separations and spatial resolution of HST ACS.
$N{\rm p0}$ is the number of star pairs in a population at zero age, which is similar for all panels.
$N{\rm obj}$ is number of objects or points in the shown ranges, and for the given populations.
Two ranges shown by blue lines include BSs that are compared to those of NGC1846 \citep{mack08}.
Errors in $V$ and $V-I$ follow Gaussian distributions with standard deviations of 0.01 and 0.014\,mag.
``nonrotation + rotation'' means half non-rotating stars and half rotational stars. }
\end{figure}

\end{document}